\documentclass[%
 reprint,
nofootinbib,
nobibnotes,
 amsmath,amssymb,
 aps,
prd,
]{revtex4-2}

\usepackage{subfigure}

\usepackage{amsmath}
\usepackage{amssymb}
\usepackage{mathtools}

\usepackage{hyperref}

\usepackage{graphicx}
\usepackage{color,xcolor}

\usepackage{ulem}

\usepackage{comment}

\graphicspath{
  {./}
  {./figures/}
}

\newcommand{\beq}{\begin{equation}}
\newcommand{\eeq}{\end{equation}}
\newcommand{\bea}{\begin{eqnarray}}
\newcommand{\eea}{\end{eqnarray}}
\newcommand{\ba}{\begin{array}}
\newcommand{\ea}{\end{array}}
\newcommand{\bit}{\begin{itemize}}
\newcommand{\eit}{\end{itemize}}

\newcommand{\eq}[1]{Eq.~(\ref{#1})}

\definecolor{purple}{rgb}{0.5,0,0.5}

\begin{document}

\title{ Double-layered vacuum bubbles and cosmological phase transitions}

\author{Dongdong~Wei$^{1,2,3}$ }
\email{weidd@ucas.ac.cn}

\author{Zong-Kuan~Guo$^{3,2,1}$}
\email{guozk@itp.ac.cn}

\author{Qiqi~Fan$^{4}$ }
\email{fanqq5@mail2.sysu.edu.cn}

\affiliation{${}^1$School of Fundamental Physics and Mathematical Sciences, Hangzhou Institute for Advanced Study, University of Chinese Academy of Sciences, Hangzhou 310024, China}

\affiliation{${}^2$School of Physical Sciences, University of Chinese Academy of Sciences, No.19A Yuquan Road, Beijing 100049, China}

\affiliation{${}^3$Institute of Theoretical Physics, 
Chinese Academy of Sciences, Beijing 100190, China}

\affiliation{${}^4$School of Physics and Astronomy, Sun Yat-sen University, Zhuhai 519082, China}
\date{\today}

\begin{abstract}
We investigate the evolution and formation of double-layered vacuum bubbles during cosmological phase transitions with multiple vacua. We employ lattice simulations to show that flyover transitions can produce double-layered vacuum bubbles by overcoming successive potential barriers, thereby suggesting a novel bubble vacuum configuration in cosmological phase transitions. The evolution of these bubbles, including wall acceleration, collisions, and the formation of trapped regions, is explored through numerical simulations. Our results show that the dynamics of double-layered bubbles differ significantly from standard single-wall bubbles, with implications for cosmological observables such as gravitational wave production and baryogenesis. 
\end{abstract}

\maketitle
\section{Introduction} 
Phase transitions in the early Universe play a crucial role in shaping its subsequent evolution and the formation of cosmological structures~\cite{Vilenkin:2000jqa,Rubakov:2017xzr}. Traditionally, such transitions have been studied in the context of quantum tunneling, where a scalar field initially trapped in a metastable vacuum can tunnel through a potential barrier to reach a lower-energy vacuum~\cite{Ellis:1990bv, Linde:1991sk}. This process provides a quantum-mechanical description for bubble nucleation, and its consequences have been widely investigated in both flat and curved spacetimes~\cite{Coleman:1980aw}. Quantum tunneling not only determines the decay rate of metastable vacua, but also be relevant for a range of cosmological phenomena, including first-order phase transitions(which may be thermal or quantum), some baryogenesis scenarios~\cite{Morrissey:2012db,Barger:2008jx,Bian:2017wfv}, and the generation of stochastic gravitational wave backgrounds~\cite{Huber:2008hg,Caprini:2007xq,Caprini:2009fx,Jinno:2017fby,Lewicki:2020jiv,Wei:2022poh}.

Beyond the standard instanton (tunneling) description, vacuum decay can also be initiated by quantum or thermal fluctuations that place the field in an over-the-barrier configuration (or generate a sufficiently large initial field velocity), after which the subsequent nonlinear evolution proceeds according to the classical equations of motion, allowing the field to roll over the barrier and complete the transition. This fluctuation-induced channel is often referred to as the stochastic or ``flyover'' picture~\cite{Braden:2018tky, Blanco-Pillado:2019xny, Tranberg:2022noe, Wang:2019hjx, Batini:2023zpi, Easther:2009ft, Giblin:2010bd}.
A different classically driven channel can also occurs in multi-vacuum scenarios, where it is typically triggered by collisions between vacuum bubbles formed through quantum tunneling. When these bubbles collide, they may generate enough energy to push the scalar field over potential barriers, leading to the formation of new bubbles in deeper vacua, without the need for additional quantum tunneling. Both mechanisms have been studied in various contexts, including stochastic methods and numerical simulations in one-dimensional models, providing insights into the decay rates, bubble dynamics, and the role of non-quantum fluctuations. 

In models with multiple metastable vacua, the landscape becomes richer, allowing for the coexistence of quantum tunneling and classical vacuum decay mechanisms. Depending on the potential structure and the physical conditions of the early Universe, a metastable vacuum can decay via one or more of these processes, leading to complex bubble configurations, such as double-layered vacuum bubbles. These multi-step or multi-channel decay processes can significantly modify the dynamics of phase transitions, influencing wall-related physics, collisions, and the resulting field configurations.

In this work, we investigate flyover vacuum decay in scalar field theories with multiple vacua, focusing on the formation, evolution, and interactions of double-layered vacuum bubbles. We adopt flyover vacuum decay with initial velocity fluctuations, and numerically simulate bubble formation, collisions, and the formation of trapped regions. Our study aims to characterize the distinct dynamics of double-layered bubbles, compare them with standard single-wall transitions, and explore their implications for cosmological observables such as gravitational waves. 
The paper is organized as follows. 
In Sec. \ref{sec:mechanism}, we introduce the mechanism for the formation of double-layered bubbles and the flyover vacuum decay. 
In Sec. \ref{sec:dynamics}, we present numerical results on the formation and evolution of double-layered bubbles. 
Sec. \ref{sec:conclusion} provides the conclusion and discussion.
\section{Mechanism for the formation of double-layered bubbles\label{sec:mechanism}}

\begin{figure}
\centering 
\hspace{2.19 cm}
\includegraphics[width=0.4\textwidth]{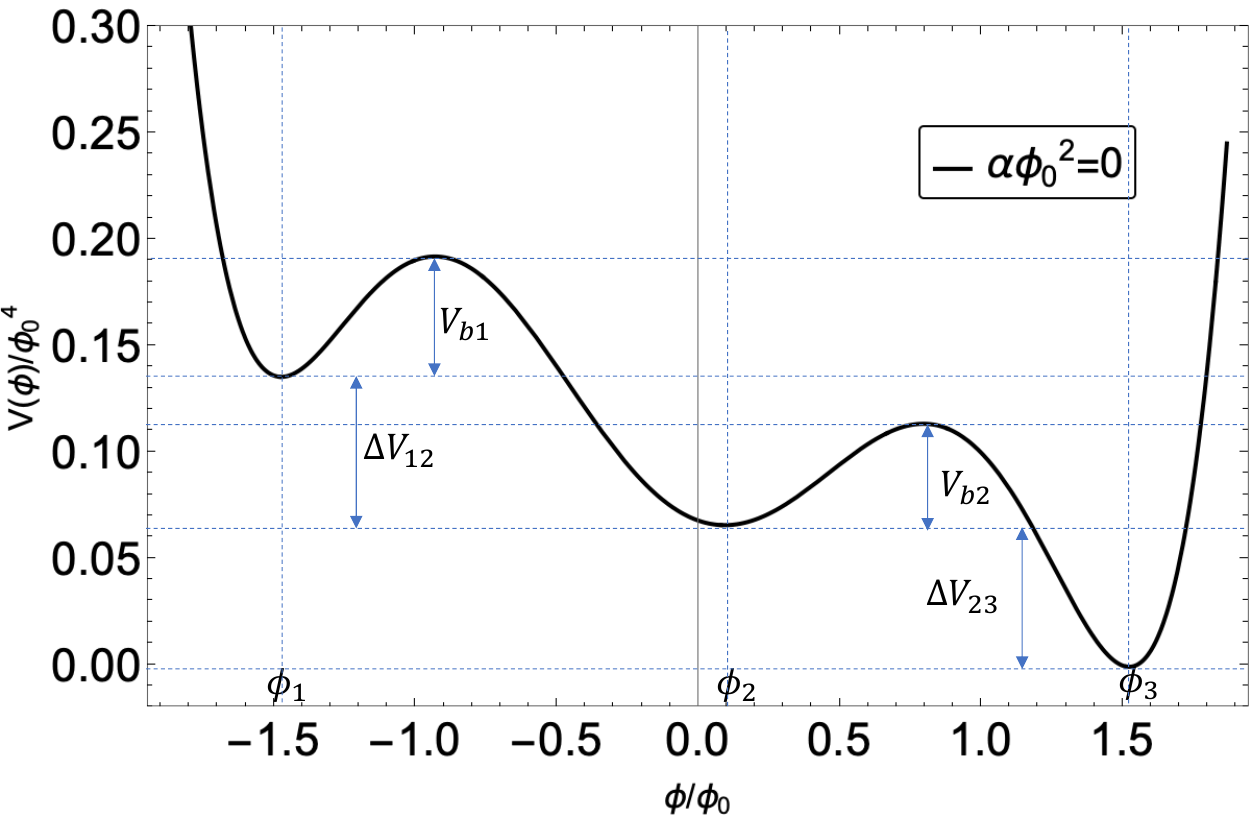}
\caption{Potential \eqref{eq:potential} with three minima at $\phi_1$, $\phi_2$ and $\phi_3$. The heights of the two potential barriers are labeled as $V_{b1} $and $V_{b2}$, and the potential differences between the adjacent minima are denoted by $\Delta V_{12}$ and $ \Delta V_{23}$. } 
\label{fig:potential2} 
\end{figure}

To investigate double-layered vacuum bubble formation, we consider vacuum decay arising from flyover vacuum decay~\cite{Blanco-Pillado:2019xny, Wang:2025eee}. We take a potential with three vacua as shown in Fig.~\ref{fig:potential2},
\begin{equation}
V(\phi)=\frac{\lambda}{4} \phi^2\left(\phi^2-\phi_0^2\right)^2+\epsilon \phi_0^5\left(\phi-\phi_0\right)+\alpha  \phi^2\phi_0^2.
\label{eq:potential}
\end{equation}
Here, the parameter $\epsilon  \phi_0^2$ must be sufficiently small to ensure the existence of potential barriers. When $\epsilon \phi_0^2 = 0$, the potential exhibits a $Z_2$ symmetry. The dimensionless parameter $\alpha$ governs the difference between $\Delta V_{12}$ and $\Delta V_{23}$, typically $\Delta V_{12} \approx \Delta V_{23}$ at $\alpha=0$.
 Such scalar potential with three distinct vacua can naturally arise in various cosmological settings, such as in the axion monodromy models~\cite{Hebecker:2016vbl}. In these scenarios, the scalar field may become trapped in a local minimum due to preceding physical processes, such as during the inflationary epoch, and subsequently transition to a lower vacuum state through thermal fluctuations or quantum tunneling. This behavior closely resembles the picture considered in our analysis.

If the scalar field $\phi$ starts off localized in a metastable minimum at $\phi = \phi_1$, the surrounding potential barrier prevents classical evolution toward lower-energy vacua, yet still permits quantum tunneling to these vacua. In real space, such tunneling events correspond to the nucleation of vacuum bubbles whose interiors reside in the $\phi = \phi_2$ or $\phi = \phi_3$ vacuum. In the potential under consideration, quantum tunneling can give rise to three distinct types of vacuum bubbles:
$\phi_1 \to \phi_2, \phi_2 \to \phi_3$, and $\phi_1 \to \phi_3$.
The tunneling rate is approximately
$\Gamma \approx e^{-S_\mathrm{E}}$,
where $S_\mathrm{E}$ is the Euclidean action~\cite{Coleman:1977py,Callan:1977pt,Linde:1981zj}
\beq
S_\mathrm{E}=\int \mathrm{d}^4x \left[ \frac{1}{2} (\partial_\tau \phi)^2 + \frac{1}{2} (\nabla \phi)^2 + V(\phi) \right],
\eeq
where $\tau$ is the Euclidean time. The dominant contribution comes from the path that minimizes $S_\mathrm{E}$, corresponding to the O(4)-symmetric bounce solution of
\beq
\partial_{r_{\mathrm{E}}}^2 \phi + \frac{3}{r_{\mathrm{E}}} \partial_{r_\mathrm{E}} \phi = V^{\prime}(\phi),
\eeq
with $r_\mathrm{E}^2 = \tau^2 + x^2 + y^2 + z^2$ and the boundary conditions $\partial_{r_\mathrm{E}} \phi = 0$ at $r_\mathrm{E} = 0 $ and $\phi \to 0$ as $r_\mathrm{E} \to \infty$.

\begin{figure}[t]
\centering 
\hspace{2.19 cm}
\includegraphics[width=0.4\textwidth]{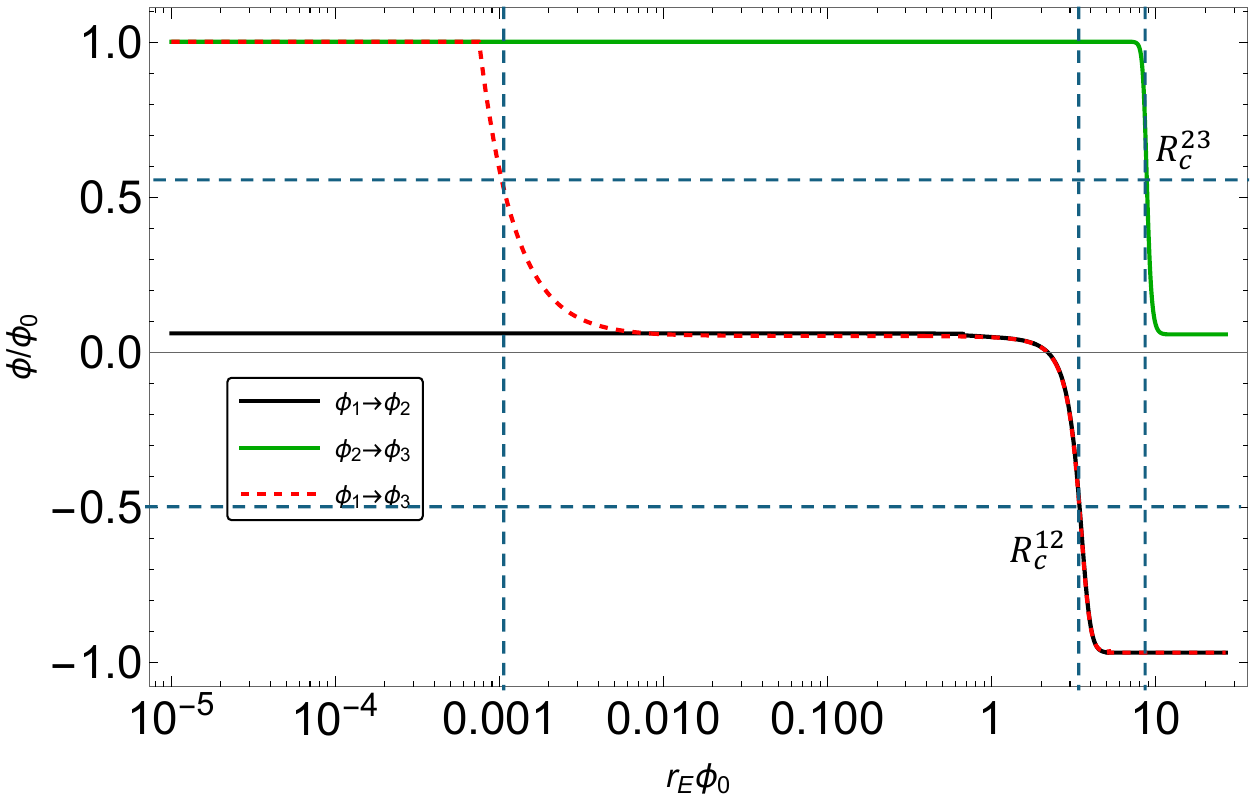}
\caption{Vacuum bubble configuration arising through a quantum tunneling process. The potential parameters are chosen as $\alpha = 0.05$, $\phi_0 = 1.5$, and $\epsilon \phi_0^2= 0.068$. Under these conditions, the scalar field tunnels from the false vacuum into a true vacuum region, forming a bubble with a characteristic profile determined by the bounce solution.} 
\label{fig:bubble_profile} 
\end{figure}
In Fig.~\ref{fig:bubble_profile}, we use FindBounce~\cite{Guada:2020xnz} to numerically obtain the critical bubble configurations $\phi_c$ for the three tunneling channels. 
For the $\phi_1 \to \phi_2$ and $\phi_2 \to \phi_3$ transitions, we find the Euclidean critical radius $R_c^{12}$ and $R_c^{23}$ as the radial positions in the numerical field profiles where the field reaches the midpoint between the two local minima.\footnote{
Here ``Euclidean critical radius'' refers to a characteristic size extracted from the $O(4)$-symmetric Euclidean bounce profile.
In the thin-wall limit one can make this notion precise by minimizing the thin-wall Euclidean action
$S_E(R)\simeq 2\pi^2 R^3\sigma-\frac{\pi^2}{2}R^4\Delta V$, which yields $R_c=3\sigma/\Delta V$.
By contrast, the Lorentzian (real-time) growth/collapse threshold obtained from the competition between surface and volume energy,
$E(R)\simeq 4\pi R^2\sigma-\frac{4\pi}{3}R^3\Delta V$, gives $R_0=2\sigma/\Delta V$.
Away from the thin-wall regime, the bubble does not have a unique ``radius''; following \cite{Cutting:2018tjt} we use the midpoint criterion as a practical and visually robust proxy for the bounce size, while the actual expansion/collapse is determined by real-time evolution.}
Explicitly, this can be written as~\cite{Cutting:2018tjt}
\beq
\phi_c\left(R_{\mathrm{c}}^{12}\right) = \frac{\phi_2 + \phi_1}{2}, \qquad
\phi_c\left(R_{\mathrm{c}}^{23}\right) = \frac{\phi_2 + \phi_3}{2}.
\eeq

In the thin-wall limit, the threshold size beyond which a nucleated bubble will grow rather than collapse is set by the Lorentzian critical radius $R_0$, which is related to the Euclidean critical radius by
$R_0 = \tfrac{2}{3} R_c$. It is interesting to note that, for the benchmark parameters chosen in this work, the tunneling rate for
$\phi_1 \to \phi_3$ is comparable to that for $\phi_1 \to \phi_2$. In Fig.~\ref{fig:bubble_profile}, the red dashed curve represents the critical bubble profile for the $\phi_1 \to \phi_3$ transition. This profile can be naturally decomposed into two parts: one corresponding to the $\phi_2 \to \phi_3$ region and the other to the $\phi_1 \to \phi_2$ region,
\beq
\phi_c=
\begin{cases}
\phi_{\mathrm{part1}}, & \phi > \phi_2, \\
\phi_{\mathrm{part2}}, & \phi \leq \phi_2 .
\end{cases}
\eeq
To determine whether different parts of the $\phi_1 \to \phi_3$ configuration can grow after nucleation, we compare their characteristic radii with the corresponding Lorentzian critical values $R_0^{12}$ and $R_0^{23}$. It can be seen that the radius of $\phi_{\mathrm{part2}}$ is equal to $R_{c}^{12}$. As a result, this part of the field configuration can successfully grow once nucleated. In contrast, the radius of $\phi_{\mathrm{part1}}$ is much smaller than $R_{0}^{23}$, causing this part of the bubble to collapse. Consequently, the $\phi_1 \to \phi_3$ tunneling process cannot be dynamically distinguished from the $\phi_1 \to \phi_2$ transition. In practice, quantum tunneling nucleation effectively produces only $\phi_1 \to \phi_2$ vacuum bubbles in this setup.

In addition to vacuum bubble nucleation via quantum tunneling, classical mechanisms can also enable the scalar field to overcome the potential barrier and facilitate the phase transition. This process, known as flyover vacuum decay, essentially involves describing vacuum decay through a classical stochastic framework, where quantum or thermal fluctuations are modeled as initial fluctuations following a Gaussian distribution for a single real scalar field in 1+1 dimensions. Unlike quantum tunneling, in this scenario the scalar field acquires a non-zero configuration that trigger the field to roll classically over the barrier, allowing bubble nucleation via a classical pathway.

To study the subsequent evolution of the scalar field under this initial condition, we solve its equation of motion in the relevant cosmological setting. We consider phase transitions that occur on timescales much shorter than the Hubble time $H_*^{-1}$ at the epoch of the transition. In this case, the effect of cosmic expansion is negligible, and the scalar field satisfies
\beq
\square \phi - \frac{\mathrm{d}V(\phi)}{\mathrm{d}\phi} = 0 ,
\eeq
with $V(\phi)$ being the scalar potential.

Regardless of the mechanism responsible for the formation of vacuum bubbles, we assume that they possess full spherical symmetry, such that the scalar field $\phi$ depends only on the radial coordinate $r$ and time $t$. Under this assumption, the standard Klein–Gordon equation in flat Minkowski spacetime takes the form
\beq
\ddot{\phi} = \phi^{{\prime}{\prime}} + \frac{2}{r} \phi^{\prime} - \frac{\mathrm{d}V(\phi)}{\mathrm{d}\phi},
\label{eom11}
\eeq
where dots and primes denote derivatives with respect to time $t$ and radius $r=\sqrt{x^2+y^2+z^2}$, respectively.	This assumption enables a significant reduction in computational cost, making numerical simulations of vacuum bubble dynamics more tractable.

Unlike approaches that model flyover vacuum decay using initial field fluctuations, we follow the prescription of Refs.~\cite{Blanco-Pillado:2019xny, Wang:2019hjx}, which instead introduces fluctuations in the velocity of the field while keeping the field itself at the local metastable vacuum value. This choice is motivated by the fact that representing the fluctuations as those of a free quantum field is more robust in this case, while still yielding results consistent with the field-fluctuation approach. Specifically, the scalar field is initially homogeneous at $\phi = \phi_1$, while its velocity profile occasionally acquires, through vacuum fluctuations, a localized Gaussian form
\beq
\dot{\phi}(t=0, r) = A \exp\left(-\frac{r^2}{2 R^2}\right),
\eeq
where $A$ denotes the amplitude of the initial velocity fluctuation, characterizing the strength of the perturbation, while $R$ represents its spatial width, determining the size of the affected region. 
For such a fluctuation to seed an expanding bubble, its width must exceed the (Lorentzian) critical expansion radius, which in the thin-wall limit can be estimated as $R_0 \equiv R_0^{12} \simeq 2\sigma^{12} / \Delta V_{12}$, where the wall tension is

\beq
\sigma^{12} = \int_{\phi_{1}}^{\phi_{2}} \sqrt{ 2 \left[ V(\phi) - V{(\phi_2)} \right] } \, \mathrm{d}\phi .
\eeq
Additionally, the amplitude $A$ must be larger than $\sqrt{2V_{b1}}$ to overcome the first potential barrier $V_{b1}$ via classical evolution.
To uniformly specify, the value of $A$ at $\dot{\phi}(t = 0, r = R)$ is defined as $A_0$ where it just equals $\sqrt{2 V_{b1}}$.

This initial condition ansatz was originally proposed by Jose J. Blanco-Pillado et al.~\cite{Blanco-Pillado:2019xny}, and recent studies~\cite{Pirvu:2023plk} have shown its consistency with numerical simulations of vacuum decay in $1+1$ dimensions starting from a metastable state. These simulations demonstrate that an initial Bose-Einstein distribution of fluctuations leads to bubble nucleation with Gaussian-distributed center-of-mass velocities, typically preceded by the formation of oscillons. 

In our simulations,to numerically solve \eq{eom11}, the $1+1$ dimensional setup uses a grid size of $12,000$, with a lattice length of $L/R_0 = 10$. In the $2+1$ dimensional setup, to numerically solve \eq{eom21}, the grid consists of 500 points in each dimension, with $L/R_0 = 20$. In our simulations, we use a second-order finite difference scheme to discretize the equations. The time evolution is advanced using the leapfrog algorithm. We choose the timestep $\Delta t$ and lattice spacing $\Delta x$ such that $\Delta t = 0.1 \Delta x$, ensuring numerical stability and accuracy. As demonstrated in Refs.~\cite{Cutting:2018tjt, Cutting:2020nla}, this choice provides excellent energy conservation properties. In the numerical calculations presented below, the potential parameters in $V(\phi)$ are fixed to the benchmark values used in Fig.~\ref{fig:bubble_profile}. We have checked numerical convergence by repeating representative runs with increased resolution, and we find that the resulting field evolutions are consistent within numerical accuracy.

\begin{figure}
\centering 
\hspace{2.19 cm}
\includegraphics[width=0.4\textwidth]{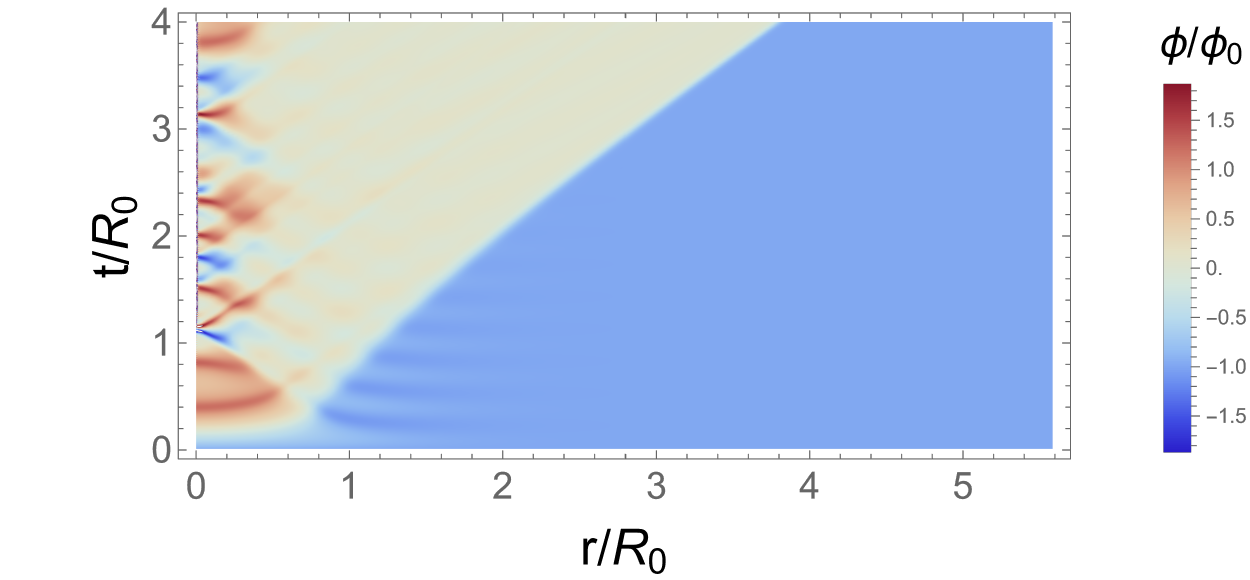}
\includegraphics[width=0.4\textwidth]{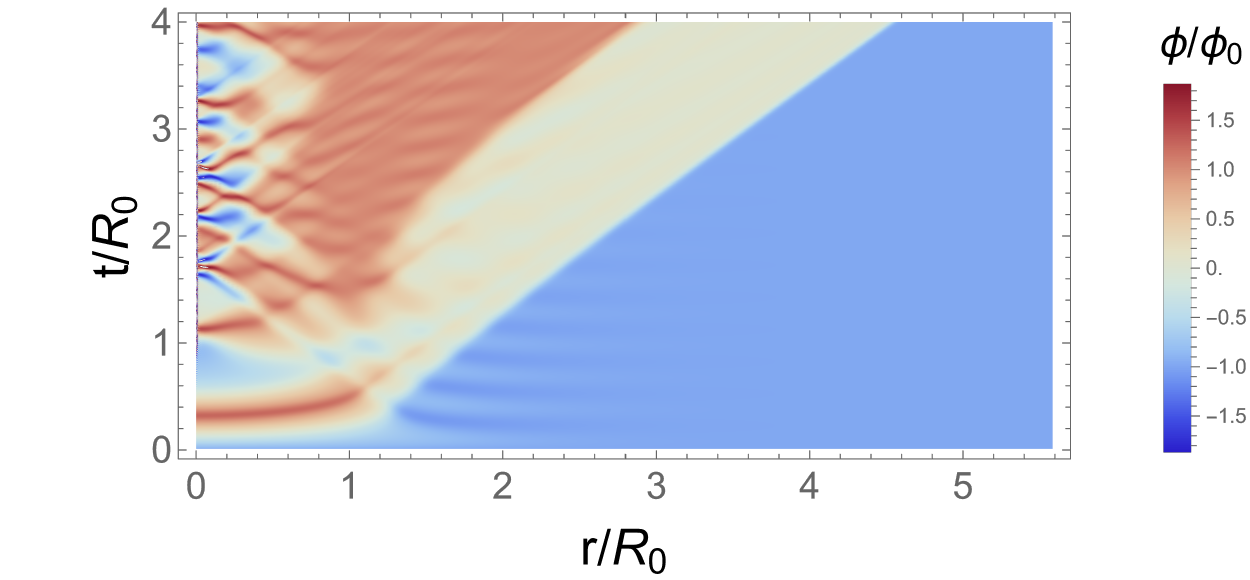}
\caption{Numerical simulation results with parameter values $A/A_0 = 1$, $R/R_0=0.8$ and $\alpha=0$ (upper), and $A/A_0 = 1.2$, $R/R_0=1.1$ and $\alpha=0$ (lower), respectively. The comparison illustrates how varying $A$ and $R$ affects the evolution of the scalar field configuration.  } 
\label{fig:double_layer_bubbles} 
\end{figure}
We consider two representative choices of the initial condition parameter with the corresponding evolution shown in Fig.~\ref{fig:double_layer_bubbles}. Vacuum transitions typically begin with the nucleation of $\phi_2$ bubbles within the $\phi_1$ background. 
For $A/A_0 = 1.2$ and $R/R_0=1.1$, the subsequent evolution toward the true vacuum $\phi_3$ --- either via $\phi_2 \to \phi_3$ or direct $\phi_1 \to \phi_3$ transitions --- depends sensitively on the relative vacuum energy gaps $\Delta V_{13}$ and $\Delta V_{23}$. 
If the $\phi_2$ vacuum is sufficiently long-lived, a two-step tunneling sequence may dominate the dynamics, driving the scalar field from the intermediate metastable state into the deeper true vacuum even when the tunneling rate to $\phi_3$ is highly suppressed. In the presence of three vacua, the flyover mechanism exhibits qualitatively distinct behavior compared to the standard two-vacuum scenario.

 When the initial velocity is small enough, the dynamics of the scalar field are essentially indistinguishable between the two-vacuum and three-vacuum potentials, as the field cannot overcome either potential barrier. However, for sufficiently large initial velocities --- particularly those with a Gaussian spatial profile --- the central region of the perturbation, where the velocity is highest, can overshoot the intermediate vacuum $\phi_2$ and reach the true vacuum $\phi_3$, while the outer regions, having lower velocities, only transition to $\phi_2$. This results in the formation of a special \textbf{ double-layered vacuum bubbles} structure, in which an inner bubble of $\phi_3$ is nested within an outer shell of $\phi_2$.

\begin{figure}
\centering 
\hspace{2.19 cm}
\includegraphics[width=0.4\textwidth]{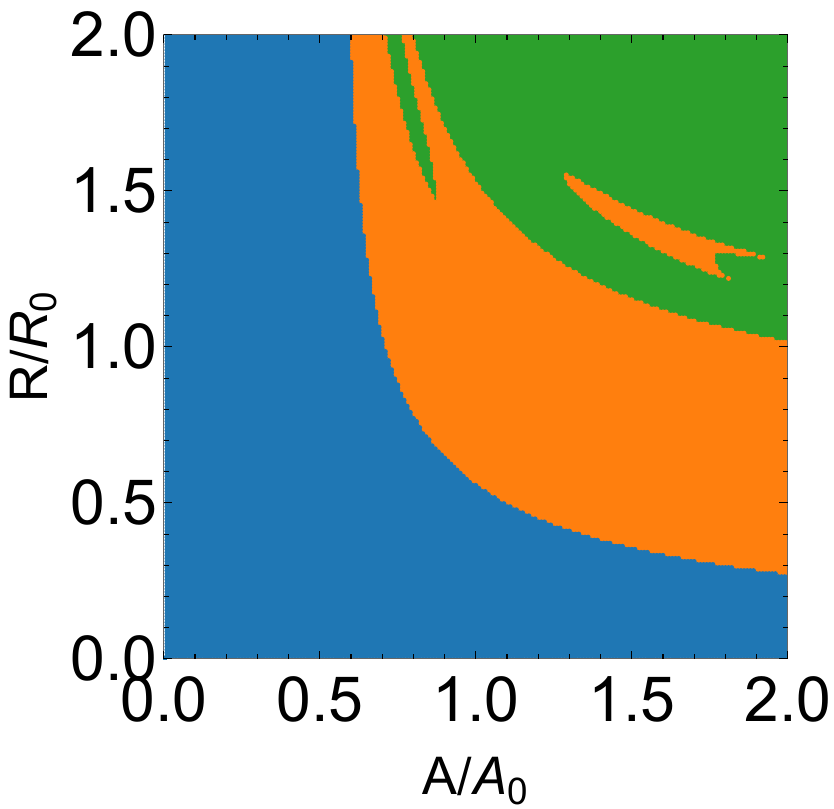}
\caption{Parameter space where the blue region indicates the absence of a phase transition, the orange region suggests a transition expanding toward the $\phi_2$ vacuum, and the green region indicates a transition to the $\phi_3$ vacuum.} 
\label{fig:para_scan} 
\end{figure}

In this scenario, double-layered vacuum bubbles are generated via flyover transitions. The occurrence of such configurations depends on the dimensionless parameters $A/A_0$ and $R/R_0$, whose values are determined by model-dependent physical quantities such as the temperature and the scalar field mass. Within the framework of our model, we treat $A/A_0$ and $R/R_0$ as free parameters and perform a systematic scan over their values, as shown in Fig.~\ref{fig:para_scan}. The nonlinear nature of the equations of motion leads to intricate structures—for instance, the appearance of a small orange region embedded within the green domain and a small green region within the orange area. Larger $A/A_0$ and $R/R_0$ correspond to greater initial kinetic energy, making it easier for the scalar field to overcome both potential barriers, whereas smaller values suppress such transitions. The resulting phase boundaries align with physical intuition and suggest the existence of critical values in both the $A/A_0$ and $R/R_0$ directions, beyond which vacuum decay can occur.

\section{Dynamics of double-layered bubbles \label{sec:dynamics}}

Having identified the parameter space for the formation of double-layered vacuum bubbles, we next investigate their stability. The stability of such configurations depends on the relative expansion velocities of the inner and outer walls: if the inner wall expands faster than the outer wall, it can catch up and collapse the double-layered structure into a single wall. To analyze the wall velocities, one can first assume the formation of a $\phi_1 \to \phi_2$ bubble, which allows for a straightforward calculation of the outer wall velocity. After a time delay, a $\phi_2 \to \phi_3$ bubble is generated inside the existing $\phi_1 \to \phi_2$ bubble, enabling the determination of the inner wall velocity. In this way, the velocities of the inner and outer walls can be treated as those of two independent single bubbles, allowing for a straightforward decomposition of the double-wall problem.

In the potential (\ref{eq:potential}), increasing the parameter $\alpha$ lowers the values of $\Delta V_{23}$, while enhancing the potential difference between $\Delta V_{12}$. Since these changes can significantly affect the post-nucleation dynamics, tracking the subsequent bubble growth during the acceleration stage requires solving the full equation of motion derived from the Lagrangian
\begin{equation}
\ddot{r}+2 \frac{1-\dot{r}^2}{r}=\frac{p}{\sigma}\left(1-\dot{r}^2\right)^{\frac{3}{2}},
\label{eom:bubble}
\end{equation}
where  $p$  is the pressure difference across the bubble wall, and  $\sigma$ represents the wall tension. This thin-wall description is expected to become increasingly accurate once the wall becomes relativistic, since the wall thickness is Lorentz-contracted in the lab frame.

\begin{figure}
\centering 
\hspace{2.19 cm}
\includegraphics[width=0.4\textwidth]{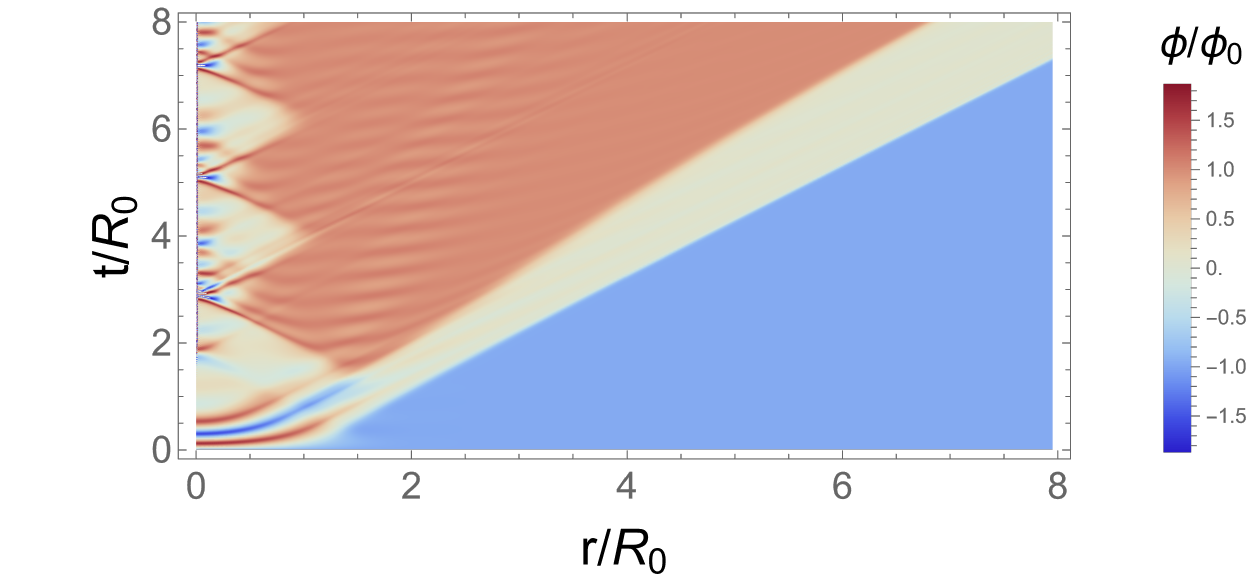}
\includegraphics[width=0.4\textwidth]{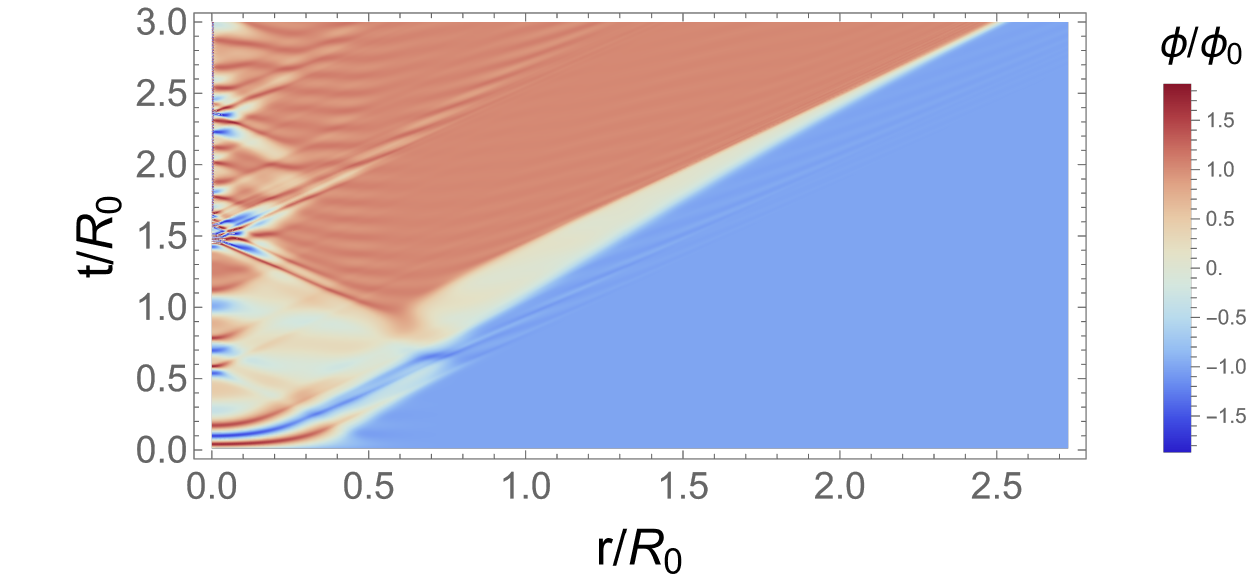}
\caption{Dynamics of vacuum decay with $\alpha = 0.035$, $\phi_0 = 1.5$, $A/A_0=7$ and $R/R_0=0.59$ (top),
and with $\alpha = -0.038$, $\phi_0 = 1.5$, $A/A_0=11$ and $R/R_0=0.2$ (lower), respectively.  } 
\label{fig:different_v} 
\end{figure}

To better understand the relationship between the wall velocity and the bubble radius, we simplify \eq{eom:bubble}. The Lorentz factor $\gamma$ characterizes the wall velocity, defined as $\gamma = 1 / \sqrt{1 - v^2/c^2}$. It can be solved analytically with the initial condition that the wall is at rest. The solution is given by~\cite{Ellis:2019oqb}
\beq
\gamma = \frac{p r}{3 \sigma} + \frac{R_c^2}{r^2} - \frac{p R_c^3}{3 \sigma r^2} \approx \frac{2 r}{3 R_c} + \frac{R_c^2}{3 r^2},
\label{motion:wall}
\eeq
where we have assumed that the initial radius is only slightly larger than the critical radius $R_c$.
As shown in Fig.~\ref{fig:different_v}, the parameter $\alpha$ directly affects the potential differences $\Delta V_{12}$ and $\Delta V_{23}$, leading to distinct evolutionary outcomes. This can be understood from \eq{motion:wall}: the velocity of a bubble wall is determined by the pressure difference across it. A larger potential difference drives the outer wall to move faster, while the inner wall propagates more slowly. In contrast, a negative $\alpha$ can cause the inner wall to overtake the outer wall, resulting in a collision and subsequent merging of the two walls. Conversely, if the outer wall expands faster than the inner wall, the vacuum bubble can stably maintain a double-wall structure.

\begin{figure}
\centering 
\hspace{2.19 cm}
\includegraphics[width=0.45\textwidth]{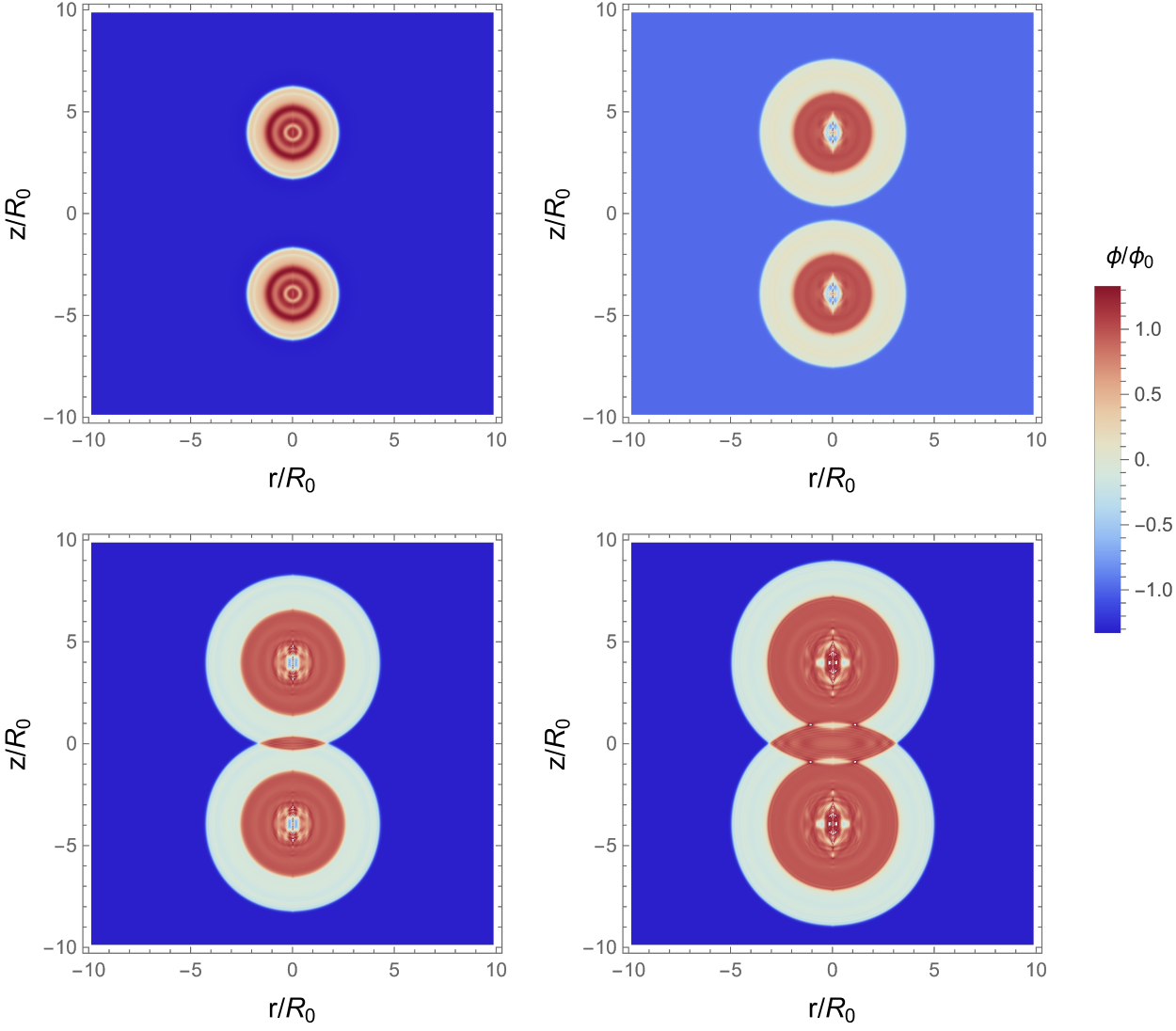}
\caption{Double-layered vacuum bubbles are formed and the collision occurs before any significant interaction between the walls. The top row corresponds to $t/R_0 = 1.56$ and $t/R_0 = 3.16$, and the bottom row corresponds to $t/R_0 = 3.95$ and $t/R_0 = 4.73$, from left to right. The simulation parameters are $A/A_0 = 1.7$, $R/R_0 = 0.98$, $D/R_0=3.95$, and $\alpha = 0$.   } 
\label{fig:collision} 
\end{figure}

Furthermore, when two double-layered bubbles collide, they can each destabilize the other's inner and outer walls. Assuming that the bubbles are aligned along the $z$-axis, the system exhibits approximate cylindrical symmetry, which allows us to simplify the analysis of their subsequent evolution. In this setup, the dynamics of the scalar field during the collision can be described by the Klein–Gordon equation in cylindrical coordinates
\beq
\partial_t^2 \phi - \partial_r^2 \phi - \frac{1}{r}\partial_r \phi - \partial_z^2 \phi = -\frac{\mathrm{d}V}{\mathrm{d}\phi}, \quad r = \sqrt{x^2+y^2}.
\label{eom21}
\eeq
We place two identical Gaussian wave packets at positions $z/R_0 = \pm D$, with $D$ chosen to be larger than the Gaussian width to prevent initial overlap or interference. As shown in Fig.~\ref{fig:collision}, double-layered vacuum bubbles are formed and the collision occurs before any significant interaction between the walls. At $t/R_0 \simeq 3.16$, the outer walls of the two bubbles begin to collide, which drives the regions near the bubble walls from the $\phi_2$ vacuum to the $\phi_3$ vacuum. Such elliptical vacuum bubbles formed through classical dynamics and the analysis of their formation have been discussed in~\cite{Easther:2009ft,Giblin:2010bd}. The key difference between the work presented here and their studies is that the colliding vacuum bubbles in this work originate from flyover vacuum decay.

The walls of these classically formed elliptical bubbles subsequently collide with the inner walls of the original bubbles, leading to the formation of so-called trapping regions~\cite{Jinno:2019bxw, Gould:2021dpm}, where trapping at the false vacuum occurs after the collisions. The formation of these trapping regions depends sensitively on the wall thickness and typically occurs in collisions involving thin-walled bubbles. Later, interactions between trapping regions and inner regions at $t/D \simeq 4.73$ further convert trapped $\phi_2$ regions into the $\phi_3$ vacuum. As time progresses, all vacua will eventually settle into the $\phi_3$ state by the scalar radiation. From our numerical results, we find that the energy--momentum tensor associated with the dynamics of double-layered bubbles can differ significantly from that of standard single-wall bubble collisions.
While we do not compute gravitational-wave spectra in this work, existing studies indicate that the characteristic frequency scale is controlled by macroscopic length scales (e.g.\ the mean bubble separation / typical bubble size at collision), whereas the ultraviolet (high-frequency) fall-off can be sensitive to microscopic wall-structure effects such as the wall thickness; therefore the additional scales introduced by the double-layer dynamics may imprint qualitatively new spectral structure (e.g.\ a shoulder/secondary feature or a modified UV slope), which we leave for dedicated GW simulations.

\section{Conclusion and discussion \label{sec:conclusion}}

In this work, we have studied flyover vacuum decay in a multi-vacuum system, demonstrating that it can produce double-layered vacuum bubbles. We have examined their stability, showing that the interplay between inner and outer wall velocities determines whether the double-layered structure is preserved or merged. Additionally, collisions between such bubbles can further disrupt the configuration. Our findings shed light on the dynamics of non-standard vacuum decay and may have implications for early-universe phase transitions and associated observable signatures.

The flyover vacuum decay differs fundamentally from conventional quantum tunneling. In tunneling-induced vacuum decay, bubbles typically nucleate sequentially in time, with their spatial locations distributed randomly. By contrast, flyover vacuum decay proceeds by overcoming successive potential barriers, leading to overlapping bubble centers and the emergence of double-layered structures. Our analysis of bubble nucleation in multi-vacuum potentials shows that the existence of stable double-layered bubbles indicates flyover transitions constitute an independent decay channel, distinct from standard quantum tunneling.

It was theoretically shown that stepwise tunneling processes can generate nested bubbles~\cite{Morais:2019fnm, Croon:2018new}, where new bubbles nucleate inside older ones as the transition proceeds. In such models, flyover vacuum decay can also occur, enhancing the number of bubbles formed at each stage. However, since the multi-vacuum structure arises from two or more fields, the flyover process cannot traverse successive barriers continuously, but it can still overcome individual barriers, thereby producing configurations that resemble nested bubbles. Unfortunately, although both mechanisms may coexist during the phase transition, their dynamical processes are indistinguishable.

Furthermore, the dynamics of double-layered bubbles differ significantly from single-wall bubbles. The expansion and collisions of these bubbles modify the scalar field evolution, which in turn can drive the evolution of coupled fluids, potentially leading to distinct cosmological signatures such as gravitational waves or baryogenesis. Additionally, the presence of two types of nucleated bubbles introduces multiple possible decay pathways, giving rise to gravitational wave power spectra that depend on a variety of parameters controlling the bubble dynamics. Overall, our results indicate that flyover transitions enrich the phenomenology of vacuum decay in multi-vacuum potentials, opening new possibilities for observational consequences in the early Universe.

\vspace*{5mm}

\section*{acknowledgements}
This work is supported in part by the National Natural Science Foundation of China  No. 12235019 and No. 12475067.

\vspace*{-2mm}

\bibliography{bubbles} 

\end{document}